\documentclass[1p]{elsarticle}
\usepackage{amsmath,amssymb}
\usepackage{graphicx}

\def\vec#1{{\boldsymbol#1}}                

\def\x{{\vec x}}

\begin{document}
\begin{frontmatter}
\title{Universal properties of the FQH state from the topological entanglement entropy and disorder effects}
\author{Na Jiang}
\author{Qi Li}
\author{Zheng Zhu}
\author{Zi-Xiang Hu}
\ead{zxhu@cqu.edu.cn}
\address{Department of Physics, Chongqing University, Chongqing, 401331, P. R. China} 

\begin{abstract}
The topological entanglement entropy (TEE) is a robust measurement of the quantum many-body state with topological order. In fractional quantum Hall (FQH) state, 
it has a connection to the quantum dimension of the state itself and its quasihole excitations from the conformal field theory (CFT) description.  
We study the entanglement entropy (EE) in the Moore-Read (MR) and Read-Rezayi (RR) FQH states. The non-Abelian quasihole excitation induces an extra correction 
of the TEE which is related to its quantum dimension. With considering the effects of the disorder, the ground state TEE is stable before 
the spectral gap closing and the level statistics seems to have significant change with a stronger disorder, which indicates a many-body localization (MBL) transition.
\end{abstract}
\end{frontmatter}
\section{Introduction}
  Fractional quantum Hall (FQH) liquids are remarkable many-electron systems that occur in two-dimensional electron gas with a perpendicular magnetic field~\cite{Tsui}. This is the most studied and first experimentally
  realized system in condensed matter physics that has the topological order~\cite{XW1995}. It is a typical strong correlated electron system with quenched kinetic energies by magnetic field which fails the application of the perturbation theory.
 On the other hand, comparing with the Landau theory of the quantum phase transition, there is no order parameter, or symmetry breaking to describe the phase transition between any two FQH states.
 Therefore, the understanding of the FQH effect has only benefited either from the numerical diagonalizing the Hamiltonian for finite size system, such as exact diagonalization~\cite{Laughlin83}, density matrix renormalization 
 group~\cite{Feiguin, jzzhao, hupla}, matrix product state~\cite{PollmannMPS, PollmannMPS1}, et.al., or using of model wavefunctions~\cite{Halperin83, jain89, moore91}. In a seminal paper of Moore and Read~\cite{moore91},  it was
 found that these model wavefunctions can be expressed as correlators of the electron operators in a conformal field theory (CFT),
 i.e., the so called conformal blocks. Although the model wavefunctions are not the exact ground state 
 wavefunctions of a realistic Hamiltonian, they are supposed to capture the universal properties of the FQH states such as fractional quasiparticle excitations and their 
 fusion relations, statistics, as well as exponents in the edge tunneling and quantum dimensions for quasiparticles. The most striking theoretically predicted properties of the 
 FQH quasiparticles is the emergence  of the Abelian or non-Abelian braiding statistics~\cite{moore91,Wilczek, RR1999}. The interchange of two Abelian quasiparticles
 adds a nontrivial phase on the wavefunction. They are named ``anyons'' since the phase is neither $\pi$ by fermions nor $2\pi$ by bosons. 
 The typical Abelian FQH states are the Laughlin series at  $\nu = 1/3, 1/5,  2/3 \cdots$. 
 However,  interchange  two non-Abelian quasiparticles results in a ground state unitary transformation in the topological degenerate Hilbert space.
 According to this, the non-Abelian FQH states have received much interests due to their potential applications in the topological quantum computation~\cite{Kitaev, Freedom, RMPNayak} recently.
 Thus far there are two most interesting examples as the candidates for the non-Abelian states which have been realized in experiments, 
 namely the FQH states on the first Landau level at $\nu=5/2$~\cite{Willett87} and $\nu=12/5$~\cite{xia04, AA2010}. For the even denominator FQH state at $\nu = 5/2$,  Moore and Read~\cite{moore91} 
 proposed a $p$-wave paired wavefunction as a candidate ground state. The nature of the $12/5$ state is still undetermined. However, the most exciting candidate of the ground state is the 
 $k = 3$ parafermion state proposed by Read and Rezayi~\cite{RR1999} which describes a condensate of three-electron clusters.
  
  For the FQH states, it has been established that there is a deep connection between the bipartite EE~\cite{Haque}, or entanglement spectrum~\cite{HF2008} and the  topological properties
  embedded in the ground state and its low-lying excitations. This connection is based on the CFT description of the FQH model wavefunctions. 
  For topological states in two dimensional systems, the bipartite EE satisfies the ``area law'' with a universal order $O(1)$ correction, 
namely the TEE~\cite{Pzanardi,Kitaev06,Levin}, i.e., $S = \alpha L - \gamma_t$ where the $L$ is the length of the boundary between two subsystems.
 For example, the bipartite FQH system can be implemented in both the momentum and real space for the two dimensional electron system. 
 The former is called the orbital cut (OC)~\cite{Haque} and the later real space cut (RC)~\cite{RSES}.  In this work we mostly use the RC since it
has a more accurate definition of the boundary in the ``area law''. The EE depends on the way of the partition the system, or $\alpha$ is not universal. 
However, the TEE $\gamma_t$ is a robust measurement 
of quantum entanglement in a topological phase. It has a connection to the total quantum dimension as $\gamma_t = \text{ln}\mathcal{D}$ and $\mathcal{D} = \sqrt{\sum_i d_i^2}$,
where $d_i$s are the quantum dimensions of each sector making up the topological field theory of the corresponding FQH states.  In fact, for a general RR state with
order-$k$ clustering and at filling fraction $\nu = \frac{k}{kM + 2}$, the total quantum dimension is $\mathcal{D}_{k,M} = \frac{\sqrt{(k+2)(kM+2)}}{2\sin[\frac{\pi}{k+2}]}$. 
Such as the Laughlin state at $\nu = 1/3$, the MR state at $\nu = 5/2$ and the RR state at $\nu = 12/5$ are corresponding to $M=1$ and  $k = 1, 2, 3$ RR states respectively. 
The TEE has an additional correction when a topological excitation, or a quasihole is created in the system, i.e.,
$\gamma_t^{qh} = \ln\mathcal{D} - \ln d_\alpha$ where $d_\alpha$ is the quantum dimension of the quasihole. 
In general, Abelian quasihole excitation has quantum dimension $d_\alpha = 1$ and $d_\alpha > 1$ for non-Abelian ones. 
Therefore, the behaviors of the EE, especially the TEE should be very different while a non-Abelian quasihole
is created, in other words, we can measure the quantum dimension of the quasihole from the shift of the EE before and after its excitation.

The discussions above are based on a clean system without introducing the effects of the disorder.
The topological properties are believed to be robust in the presence of a weak disorder.  When the strength
of the disorder is comparable to that of the interaction between electrons, the FQH will eventually be destroyed and the system enters into a localized 
insulating phase. The topological Chern number calculation~\cite{DNXW03} shows that the destruction of the FQHE is related to the continuous collapse of the 
mobility gap of the system. On the other hand, because of the strong electron-electron interaction in the FQH system, the localized phase driven by the disorder can be treated as 
a many-body localized phase. The transition between the ergodic and many-body localized (MBL) phase in the disordered interacting system is a 
subject of much interest recently~\cite{MSJE16, APalDA10, ScottManybody,NRRN16}.
From the prediction of the random matrix theory, the MBL phase transition results in the varying of the spectral statistics between the Poisson and
the Wigner-Dyson distribution. Recently Serbyn and Moore~\cite{MSJE16} found that the existence of the intermediate statistics between them which can be described by
 a general distribution with two parameters, $P(x, \beta, \gamma_p) = \alpha x^\beta e^{-\eta x^{\gamma_p}}$. In this work, we study the stability of the Laughlin state and 
 the change of the spectral statistics as varying the strength of the disorder.
 
  The rest of this paper is arranged as follows.
  The model and methods are introduced in Sec.\uppercase\expandafter{\romannumeral2}.
  In Sec.\uppercase\expandafter{\romannumeral3}, we consider non-Abelian nature for the MR state, RR state and corresponding 
  quasihole excitations via TEE in the clean systems. The effects of the disorder and energy statistics are discussed in Sec.\uppercase\expandafter{\romannumeral4} and 
  Sec.\uppercase\expandafter{\romannumeral5} gives the summary and discussion.

 \section{Model and methods}
  \begin{figure}
  \centering
  \includegraphics[width=7cm]{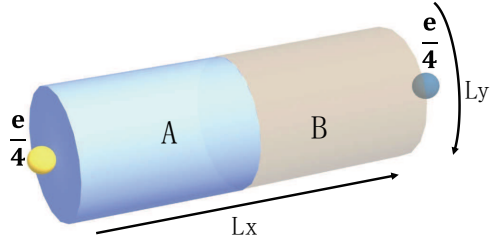}
  \caption{\label{cylinderfigure} The sketch map of the model. A cylinder is bipartited into two subsystems A and B in real-space. 
  For non-Abelian FQH state, such as the MR state at $\nu = 5/2$, one $e/4$ quasihole locates on each edge of the cylinder.}
 \end{figure}
 
 Our model is depicted as in Fig.~\ref{cylinderfigure}. Electrons are put on a cylinder with circumference $L_y$ in $y$ direction in a magnetic field perpendicular to the surface, 
 the single electron wave function in the lowest Landau level is
\begin{equation}                                                                                                                                  
\psi_{j}(\vec{r}) = \frac{1}{\sqrt{\pi^{1/2}L_y l_B}}e^{ik_{y}y}e^{-\frac{1}{2l_B^2}(x+k_{y}l_B^2)^{2}}                                                             
\end{equation}    
in which $k_{y}=\dfrac{2\pi}{L_y}j $, $ j=0,\pm1,\pm2\cdots$  are the translational momentum in the $y$ direction and the magnetic length is defined as $l_B = \sqrt{\hbar c/eB}$.  
For a finite size system, the number of orbits $N_{orb}$ equals to 
the number of the magnetic flux quantum.  Each orbit occupies an area $2\pi l_B^2$. Therefore,
the length in $x$ direction for a finite system is fixed with a given aspect ratio $\gamma$, namely $L_x/l_B = \sqrt{N_{orb} 2\pi /\gamma}$.  
 The advantage of using the cylinder geometry  is that there is no curvatures on the surface and the length $L_x$ is linearly with aspect to the system size which can be tuned easily. 
The EE is defined as $S_A = -\text{Tr}[\rho_A\ln \rho_A]$ where $\rho_A = \text{Tr}_B \rho$ is the reduced density matrix of the subsystem.  
If we make a cut in real space along $y$ direction, saying at the position $x = l_x$, the electron operators in momentum space can
be wrote as a summation of two parts:
\begin{equation}
c_m = \alpha_m a_m + \beta_m b_m
\end{equation}
where $a_m$ and $b_m$ are the operators in A/B subblock respectively. $\alpha^2_m$($\beta^2_m)$ is the probability for an electron at the $m$'th orbit locating 
in block A(B).  Therefore, 
\begin{eqnarray} \label{rcprob}
 \alpha_m^2 = \int_0^{L_y} dy \int_{-\infty}^{l_x} dx |\psi_m|^2 
 = 1 - \frac{1}{2} \text{Erfc}(l_x - \frac{2\pi}{L_y}m)  
\end{eqnarray}
and $\alpha_m^2+\beta_m^2 = 1$. For the real space partition of the many-body wavefunction, because the total particle number and translational momentum 
along $y$ direction are good quantum numbers in the subsystem, it is actually the same
as that of the particle partition with the above probabilities.  
On the other hand, the many-body model wavefunction can be generally obtained from exact diagonalizing
a model Hamiltonian with hard-core interaction. For example, the Laughlin, MR and RR wavefunctions are the most compact zero energy eigenstates for the 
two, three, and four body hard-core Hamiltonian. In the second quantized form, a $n$-body hard-core Hamiltonian in an infinite plane is 
\begin{eqnarray}
H_n &=& \sum_{\{m_i\}} V(m_1,\cdots, m_n)V(m_{n+1},\cdots, m_{2n}) c_{m_1}^+c_{m_2}^+\cdots c_{m_n}^+ c_{m_{n+1}}c_{m_{n+2}} \cdots c_{m_{2n}}
\end{eqnarray}
where the matrix elements for two, three and four-body interaction are following:
\begin{eqnarray}
&&V(m_1,m_2)  =  \sqrt{\frac{(M-1)!}{2^Mm_1!m_2!}}(m_1-m_2)  \nonumber \\
&&V(m_1,m_2,m_3) = \sqrt{\frac{(M-3)!}{4 \times 3^{M-2} m_1!m_2!m_3!}}\mathcal{A}(m_1, m_2)   \\
&&V(m_1,m_2,m_3,m_4) = \frac{1}{3} \sqrt{\frac{(M-6)!}{2^{2M-5}m_1!m_2!m_3!m_4!}}       
\sum_{\alpha < \beta < \gamma} m_\alpha m_\beta m_\gamma \mathcal{A} (m_\alpha-1, m_\beta-1)    \nonumber 
\end{eqnarray}
in which $\mathcal{A} (m_\alpha, m_\beta) = \mathcal{A}(m_\alpha(m_\alpha - 1) m_\beta) $ is the antisymmetrizer~\cite{xinprb08}. Another way of producing the 
model wavefunctions is by the help of the recursive relation of the Jack polynomials (Jacks)~\cite{bernevig08, bernevig08a, bernevig09}. Jacks are homogeneous symmetric 
polynomials specified by a rational parameter $\alpha$ and a root configuration. They satisfy a number of differential equations~\cite{Feigin02} and 
exhibit clustering properties~\cite{bernevig08,Feigin02}. For example, Jacks is one of the polynomial solutions for Calogero-Sutherland Hamiltonian:
\begin{eqnarray}\label{CSmodel}
 H_{CS}^\alpha (\{ z_i \})  = \sum_i (z_i\frac{\partial}{\partial_i})^2 + \frac{1}{\alpha}\sum_{i<j} \frac{z_i + z_j}{z_i - z_j}(z_i\frac{\partial}{\partial_i} - z_j \frac{\partial}{\partial_j}).
\end{eqnarray}
It was found~\cite{bernevig08, bernevig08a} that the FQH model wavefunctions for RR $Z_k$-parafermion states can be exactly calculated according to Eq.(\ref{CSmodel}) with 
a negative parameter $\alpha$ and a root configuration 
(or partition). The choice of the  root configuration satisfies $(k, r)$ admissibility which means there can be at most $k$ particles in $r$ consecutive orbits. 
The parameter $\alpha$  is $ -(k+1)/(r-1)$ and the corresponding filling factor $\nu = k/r$ for bosonic system ($\nu = k/(k+r)$ for fermionic system, the difference between the fermionic and bosonic wavefunction is just a Vandermonde determinant). 
For example, the Jack with $k = 2, r = 2$ ($\alpha = -3$) is the MR wavefunction at $\nu = 1$, which has root ``$20202\cdots$'' in bosonic case and $\nu = 1/2$, root ``$1100110011\cdots$'' in fermionic case. 
In this paper, we use the Jacks to produce the model wavefunctions on cylinder and bipartite it in real space. 
The EE is obtained by a summation of the entropy for all the quantum numbers in the subsystem.

\section{TEE in the non-Abelian FQH and quasihole states} 
\begin{figure}
\centering
\includegraphics[width=8cm]{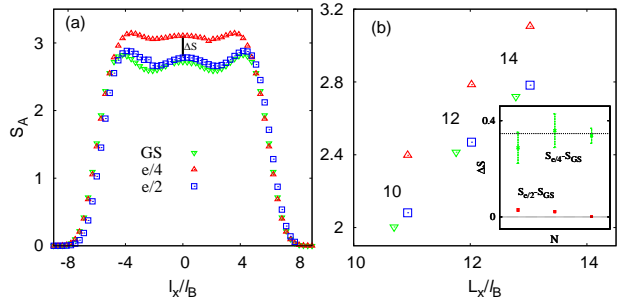}
\caption{ (a) EE ${S}_{A}$ as a function of $l_x$ for the ground state, $e/4$ and $e/2$ quasihole states of the MR state with $14$ electrons.
(b) The ${S}_{A}$ of the three states with equal partition for $10-14$ electrons. The inserted plot shows the entropy difference $\Delta S_{e/2} = S_{e/2} - S_{GS}$ and $\Delta S_{e/4} = S_{e/4} - S_{GS}$.
The horizontal lines are their corresponding expected values from CFT, i.e., $\Delta S_{e/2} = 0$ and $\Delta S_{e/4} = \log \sqrt{2} = 0.346$. The error bars are determined based on varying $L_{y}$. \label{MREE}}
\end{figure}

The MR state, or its particle-hole conjugate, is the most possible candidate trial wavefunction for the ground state of the FQH at $\nu = 5/2$.
It supports not only the Abelian quasihole with charge $e/2$ as that in the Laughlin state, but also the non-Abelian quasihole excitation with charge $e/4$. 
The origin of the non-Abelian nature of the $e/4$ quasihole excitation is the majorana zero mode embedded in the quasihole vortex. In the language of the Jacks,
the root for the ground state and $e/2$ quaihole state  are ``$11001100\cdots110011$''  and  ``$011001100\cdots110011$'' respectively. Because of the majorana nature of the 
$e/4$ excitation, it should appears in pairs in the system. For example, the root ``$101010101\cdots1010101$'' describes a configuration with one $e/4$ quasihole on each edge of the cylinder.

Fig.~\ref{MREE}(a) shows the EE as a function of the cut position $l_x$ for the above three states. Obviously, the EEs for the ground state and the 
Abelian quasihole state have neglectable discrepancy and the non-Abelian quasihole state has a larger EE in the bulk. In Fig.~\ref{MREE}(b) we plot the value of the EE
for the three states at $l_x = 0$ for different system sizes at aspect ratio $\gamma = 1.0$. Because of the Abelian nature, the difference between the $S_A^{GS}$ and $S_A^{e/2}$
becomes small while increasing the system size and approaches to zero in the thermodynamic limit. The interesting property is the increment of the EE
in the $e/4$ non-Abelian state which gradually approaches to a constant as increasing the system size. 
Up to 14 electrons, the increment approaches to $\Delta S \simeq 0.34$ which is close to the theoretical expected value $\ln d_{e/4}$ for 
the non-Abelian quasihole with quantum dimension $d_{e/4} = \sqrt{2}$.

\begin{figure}
\centering
\includegraphics[width=6cm]{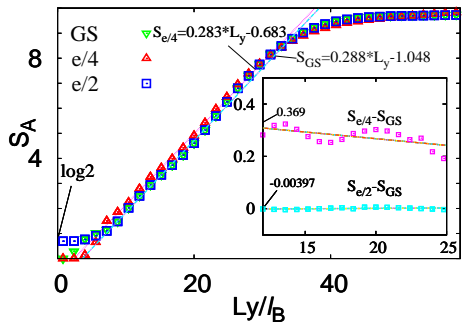}
\centering
\caption{EE ${S}_{A}$ for the ground state, $e/4$ and $e/2$ quasihole states as a function 
of $L_{y}$ for a $14$-electron system with equal partition. The data in range $L_y/l_B \in [10,25]$ are fitted by area law relation and 
the $\gamma_t^{GS} = 1.048$  and $\gamma_t^{e/4} = 0.683$ are consistent to their theoretical values. The inserted figure shows the entropy differences $S_{e/4} -S_{GS}$, $S_{e/2} - S_{GS}$ as a function of $L_y$ in the same range and the value in the thermodynamic limit are 0.369,-0.00397 which are close to their theoretical predicted values. 
\label{Fig3}}
\end{figure}

Another aspect of exploring the TEE is varying the aspect ratio of the cylinder, or the length of the cut in $y$ direction $L_y$~\cite{liqi}. For a 14-electron 
MR state with a cut at the center, we plot EE as a function of $L_y/l_B = \sqrt{N_{orb} 2\pi \gamma}$ for the above three states in Fig.~\ref{Fig3}.
While $L_y \rightarrow 0$, the system enters into the Tao-Thouless limit~\cite{tao1983, thouless84} in which the EE either goes to zero or to the 
classical Von Neumann entropy $S_{TT} = \text{ln}(2) \simeq 0.693147$ depending on whether there is an electron wave package on the cut. 
On the other hand, while $L_y$ is large, or equally $L_x$ is smaller than the 
length scale at which the two edges of the cylinder have correlations~\cite{liqi}, the area law of EE is broken down also. 
Therefore, for the system with 14 electrons as shown in Fig.~\ref{Fig3} we use the data in the medium range $L_y/l_B \in [10,25]$ and fit them by the 
formula of the area law. The EE for the ground state and $e/2$ quasihole state in this region can be fitted by $S_{GS} = 0.288L_y - 1.048$
and that for $e/4$ quasihole state can be fitted by $S_{e/4} = 0.28L_y - 0.683$. These results are consistent to the CFT prediction that the 
TEE for MR state is $\gamma_t^{GS} = \ln\sqrt{8} \simeq 1.0397$ and  $\gamma_t^{e/4} = \ln\sqrt{8} - \ln\sqrt{2} \simeq 0.693$ for $e/4$ quasihole state.
The entropy difference in the bulk,  i.e., the quantum dimensions for $e/2$ and $e/4$ quasiholes also appear as a function of $L_y$ which is shown in the inserted plot of Fig.~\ref{Fig3}.

\begin{figure}
\centering
\includegraphics[width=8cm]{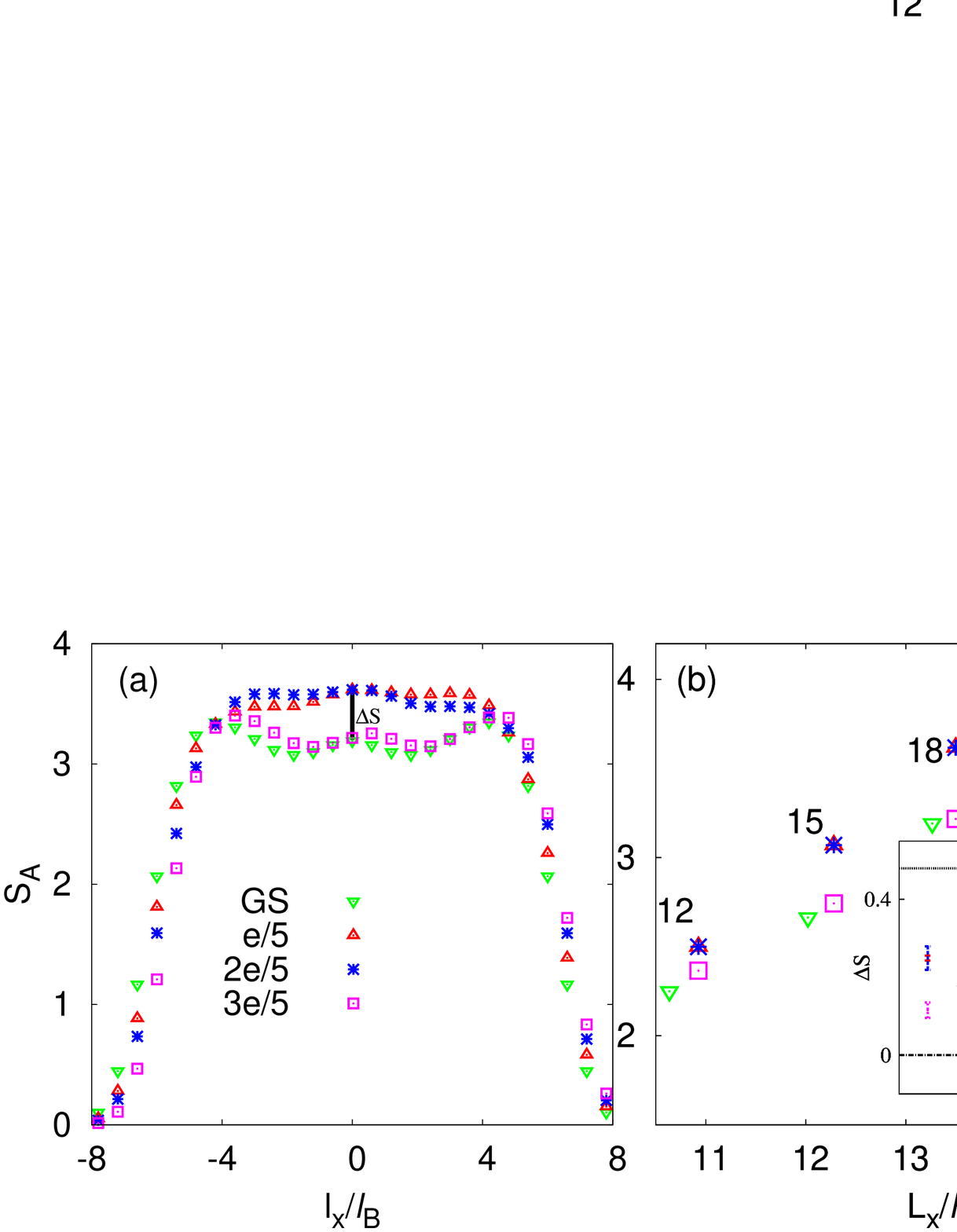}
\caption{(a) EE ${S}_{A}$ as a function of ${l}_{x}$ for the ground state, $e/5$, $2e/5$ and $3e/5$ respectively for 18-electron system. (b) The ${S}_{A}$ of the four states with two equal subsystems for $N=12-21$ systems. Insert plot: the entropy difference between ground state and
quasihole state in the bulk for different system sizes. The horizontal lines are their theoretical predicted values from CFT where $\Delta S_{1(2)e/5} - S_{GS} = \ln(\frac{1+\sqrt{5}}{2}) \simeq 0.48$
and $\Delta S_{3e/5} - S_{GS} =  0 $.\label{Fig4}}
\end{figure}

Besides the MR state, another FQH state that was proposed to support the non-Abelian quasihole excitation is the $k=3$ RR state. It is expected to be observed at 
filling fraction $\nu = 13/5$ or its particle-hole conjugate at $\nu = 12/5$.  The non-Abelian quasihole excitation in this state is called Fibonacci anyon which supports 
the universal topological quantum computation~\cite{Kitaev, Freedom, RMPNayak}. Comparing with the MR state, the RR state has smaller energy gap and higher experimental requirements. 
The exact plateau in the Hall conductance measurement so far has not been observed and the exact nature is still undetermined.
Although there are other candidates for this filling, such
as hierarchy state~\cite{Haldane83, Halperin84}, Jain's composite-fermion state~\cite{Jainbook}, Bonderson-Slingerland state~\cite{Bonderson08, Bonderson12}, 
and a bipartite composite-fermion state~\cite{wojs11, wojs13}, recent DMRG calculation~\cite{wzhu} supports that the $k=3$ RR state has lower energy and much more likely to describe
the ground state. Here we still write down the wavefunction of the RR state by Jacks and consider the properties of the quasihole excitations in the aspect of the EE.
For $k=3$ RR state, the root configurations ``$1110011100\cdots11100111$'', ``$1101011010\cdots110101101$'', ``$101101011\cdots101101011$'' and ``$01110011100\cdots11100111$'' 
correspond to the ground state, $e/5$, $2e/5$ and $3e/5$ quasihole states respectively. The $e/5$ and $2e/5$ quasihole are non-Abelian and $3e/5$ state is Abelian which is just adding a flux
at the left edge of the cylinder. The non-Abelian nature of $e/5$ and $2e/5$ can be observed in the EE as shown in Fig.\ref{Fig4}.
Although the system size is limited by the calculation of the eigenvalues of the reduced density matrix for subsystem by single value decomposition, for system with 18 electrons, it it clear that
the EE as a function of the cut position $l_x$ before and after the inserting $3e/5$ quasihole is almost invariant.
The EE for $e/5$ and $2e/5$ states are the same and have a larger value than that of the ground state.  The EEs for different states
at the center of the cylinder and their differences are shown in Fig.\ref{Fig4} (b) and inserted plot. As increasing the system size, the entropy difference for Abelian quasihole
$\Delta S = S_{3e/5} - S_{GS}$ approaches to zero and  the entropy contributions from the non-Abelian quasiholes with charge $e/5$ and $2e/5$ are getting close to the theoretical expected value 
$\Delta S = S_{1(2)e/5} - S_{GS} = \ln((1+\sqrt{5})/2) \simeq 0.48$ which is labelled by horizontal line in the plot.

\section{The disorder effects}   
\begin{figure}
\centering
\includegraphics[width=8cm]{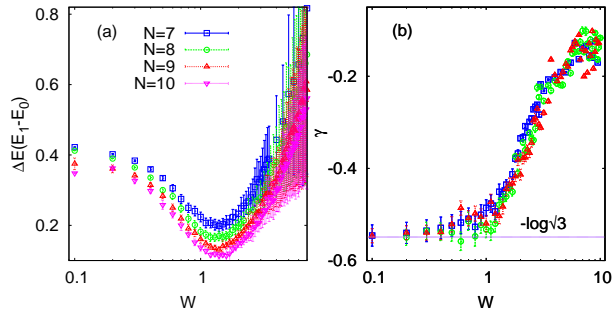}
\caption{(a) The energy gap $\Delta E = E_1 - E_0$ in the subspace $M_{tot} = 3N(N-1)/2$ as a function of the disorder strength. The minimum reaches at $W_c \simeq 1.3$ which
means the spectral gap closing. (b) The averaged TEE versus the disorder strength. The line depicts the expected value $\gamma  = -\log\sqrt{3}$ in the pure case.
\label{Fig5}}
\end{figure}

The above calculations are done in a clean system. However, in a realistic sample, the disorder is inevitable existence. 
It is known that the effect of the disorder is essential in the quantum Hall physics, such as the formation of the Hall conductivity plateaus and the transition between them.
Generally, the ground state topological properties are robust in the presence of a weak disorder. However, while the strength of the disorder is enhanced,
the ground state can be destroyed and driven into a localized Anderson insulating phase~\cite{Anderson}.  In the free case, the system is described by
a many-body interaction Hamiltonian, and therefore, the inducing of the disorder is related to the MBL physics~\cite{APalDA10}.

Recently Geraedts et.al.~\cite{ScottManybody} found that the entanglement spectrum (ES) of a spin system shows level repulsion and follows a 
semi-Poisson distribution in the MBL phase. Ref.~\cite{Zhaoliu} found that the MBL behavior is also shown
in the ES of the FQH states by introducing the disorder effect. In the following, we want to look at that to what extend, the TEE can survive with increasing the disorder strength, and we
also discuss the relation among the disorder effects and the ground state gap closing, and the MBL behavior in the energy spectrum statistics.
In the disk geometry, for simplicity, we consider the disorder effects in the model Hamiltonian with hard-core interaction.
In principle, the system breaks the translational symmetry after introducing the disorder. However, in the disorder problem, we
always need to do the sample average and the translational symmetry recovers when the disorder is randomly distributed and the number of the disorder is large enough. Here, for simplicity,
we consider an uncorrelated random potential $H_{D}=\sum _{m}U_{m}c_{m}^{+}c_{m}$~\cite{HuIJMP}  and assume that the disorder effects
are averaged firstly in each Landau orbit and then the Hamiltonian still has rotational symmetry. The $U_{m}$ denotes the averaged random 
potential on the $m$th orbit whose value is randomly chosen in the interval of $[-W/2,W/2]$. For a $N$-electron Laughlin state, it has total angular momentum
$M_{tot} = 3N(N-1)/2$. Because of the conserved rotational symmetry assumed above, we work in the same angular momentum subspace as that for
the Laughlin state. The ground state phase transition can be understood by energy level crossing between the lowest energy state and the first excited state while increasing
the strength of the disorder. We define the spectral gap as their energy difference $\Delta E = E_1 - E_0$. Its averaged value as a function of
the disorder strength is depicted in Fig.~\ref{Fig5}(a). The number of the average samples is 1000 for 10 electrons and larger for smaller systems. 
It shows that the $\Delta E$ reaches a minimum at around $W_c \simeq 1.3$ for different systems and the minimal gap at $W_c$ decreases
as increasing the system size. Finite size effect shows that $\Delta E$ drops to zero in the thermodynamic limit which means the $W_c$ is the critical
disorder strength of the ground state phase transition. Therefore, we expect that the topological properties, such as
the TEE, keeps invariant in weak disorder regions $W < W_c$ which is shown in Fig.~\ref{Fig5}(b).   
It shows that the averaged TEE stays around the expected value $\gamma  = -\log\sqrt{3}$ in the weak disorder region and dramatically increases to zero while $W > W_c$
which demonstrates that the topological properties are immune from the weak disorder.

\begin{figure}
\centering
\includegraphics[width=10cm]{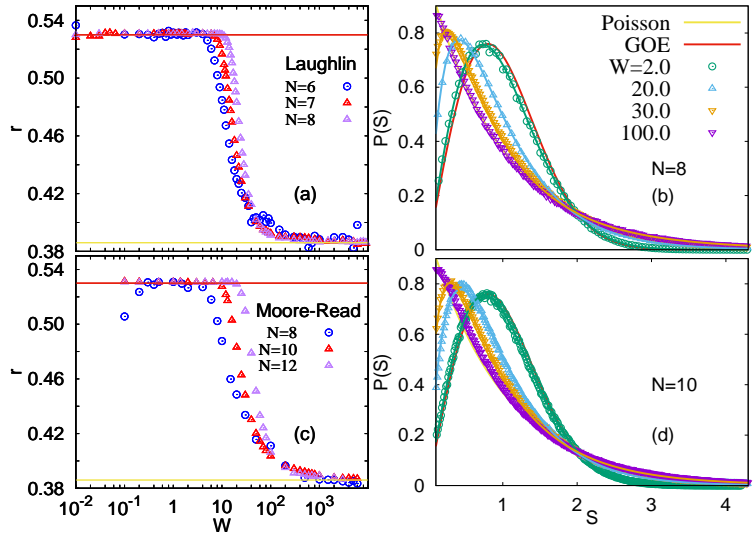}
\caption{ Evolution of the averaged ratio $r$ of the adjacent gaps as a function of the disorder strength for the model Hamiltonian of the Laughlin state (a)
and the Moore-Read state (c) and  the corresponding energy level spacing distributions for different disorder strengthes (b) (d).
\label{Fig6}}
\end{figure}

Excepting the ground state phase transition,  strong disorder in a many-body Hamiltonian can induce a MBL phase~\cite{Basko, Gornyi} in which all
of the states are insulating and their energy levels obey a Poisson distribution.  
To probe the MBL transition, it is easy to compute the ratio of adjacent energy gaps. For a sorted spectrum $\{\lambda_n; \lambda_n \le \lambda_{n+1}\}$, it is defined as
\begin{eqnarray}
 r_n = \frac{\text{min}(\lambda_n - \lambda_{n-1}, \lambda_{n+1} - \lambda_{n})}{\text{max}(\lambda_n - \lambda_{n-1}, \lambda_{n+1} - \lambda_{n})}.
\end{eqnarray}
The averaged value $r$ of the adjacent gaps is $r \simeq 0.53$ ~\cite{Alessio} for GOE ensembles,  $r \simeq 0.60$ ~\cite{Alessio} for GUE and $r \simeq 0.386$ 
for Poisson ensembles~\cite{APalDA10}. Fig.~\ref{Fig6}(a) shows the average ratio of the adjacent gaps as a function of the disorder strength.
It clearly shows that the energy levels satisfy the GOE distribution for weak disorder and evolve into Poisson for strong disorder limit. 
As a comparison, the disorder effects on the three-body model Hamiltonian, the ground state of which is the Moore-Read state, are considered in Fig.~\ref{Fig6}(c)  and (d).
The results are quit similar to that of the Laughlin state which demonstrates that the universal properties of the topological quantum state in the presence of the disorder.

\begin{figure}
\centering
\includegraphics[width=8cm]{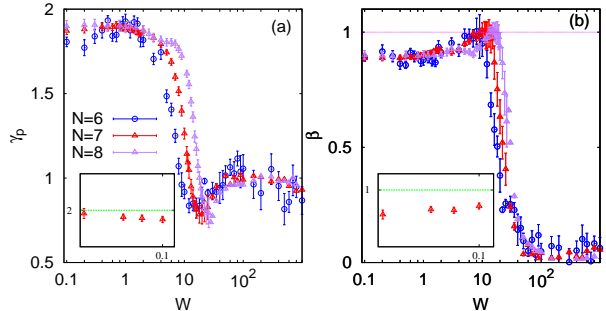}
\caption{ Evolution of the fitted values of $\gamma_{p}$($\beta$) which from the expression $\alpha x^{\beta}e^{-\eta x^{\gamma_{p}}}$ change from about 
$2.0~1.0$($1.0~0.0$) as we tune the disorder strength for $5,6,7$ electrons systems respectively. We observe the transition from GOE to Poisson. 
The inset plot is for the weak enough strength and the theoretical value can reached in the thermodynamic limit.
The fit is to an ES averaged over states in the middle $50\%$ of the full spectrum while the error averaged from the disorder.
\label{Fig7}}
\end{figure}

From Fig.\ref{Fig6}(a), we observe that the GOE distribution still persist after the spectral gap closing and there is a large region
of $W$ in which the spectral distribution obeys neither GOE nor Poisson distribution. We perform a full diagonalization 
and use the $50\%$ central part of the spectrum to investigate the energy spectral statistics. The level statistics is the distribution of the level spacing 
\begin{equation}
S\equiv \frac{\lambda _{n+1}-\lambda _{n}}{\left \langle \lambda _{n+1}-\lambda _{n}\right \rangle}.
\end{equation}
 Recently, Serbyn and Moore~\cite{MSJE16} proposed that the spectral statistics across the MBL transition can be written as a form of a 
 semi-Poisson distribution:
\begin{eqnarray} \label{fitting}
P(x, \beta, \gamma_p)  \sim  x^{\beta}e^{-\eta x^{\gamma_{p} }},
\end{eqnarray} 
in which $1\leq\gamma_p\leq2$ controls the tails of the statistics and level rigidity, and $0\leq\beta\leq1$ determines the level repulsion.
The case of $\gamma_p = 2,\beta=1$ corresponds to the Wigner-Dyson distribution for GOE ensembles and $\gamma_{p}=1, \beta=0$ corresponds to a Poisson distribution in a 
MBL phase~\cite{APalDA10}. While $\gamma_p \rightarrow 1$, it evolves to a semi-Poisson distribution with generic $\beta$ which describes an intermediate statistics between
ergodic Wigner-Dyson and MBL Poisson. Fig.~\ref{Fig6}(b) and (d) depict the level distributions with different disorder strengths (W=2.0, 20.0, 30.0, 100.0) in the model Hamiltonian. 
As a comparison, the GOE and Poisson distributions are also
plotted. It is obviously that the level distributions evolve from the GOE to Poisson with $W$ increasing. 
By using the fitting equation in Eq.~\ref{fitting}, the two optimal parameters $\gamma_{p},\beta$ for different systems are shown in Fig.~\ref{Fig7} 
as a function of the $W$.  The inserted plots show the results for much weaker disorder strengths. 
It is shown that in the limit $W \rightarrow 0$, the level distribution 
satisfies the Wigner-Dyson condition with $\gamma_p \sim 2$ and $\beta \sim 1$ which consistent to the results in Fig.~\ref{Fig6}(a). 
And in the strong disorder limit, the parameters are roughly $\gamma_p \sim 1$ and $\beta \sim 0$ which corresponds to the 
random Poisson distribution. The behavior of the $\gamma_p$ as varying $W$ is quite similar to the ratio of the adjacent gaps as shown in Fig.~\ref{Fig6}(a). 
It has a dramatic decay from $\gamma_p = 2$ to $\gamma_p = 1$ around $W \sim 10$. At the same time, the $\beta$ drops from $1$ to $0$ in this region. 
Therefore, the results of the Fig.~\ref{Fig7} are exactly consistent with that of the Fig.~\ref{Fig6} and we conclude that
the intermediate phase indeed obeys the semi-Poisson distribution. 
 
\section{Summary and discussion}
In conclusion, we have presented a systematical study of the EE for FQH states and the corrections of their 
quasihole excitations. The most important feature of the EE for FQH state is its
universal correction in the relation of ``area law'', namely the TEE which is related to the quantum dimensions of the ground 
state. Moreover, the non-Abelian quasihole excitation in the subsystem can have an extra correction in the TEE 
which is related to the quantum dimension of the quasihole itself. 
According to the behaviors of the EE  for the ground state and quasihole states, we extrapolate the 
quantum dimensions of the Abelian and non-Abelian quasiholes. 
Our results are consistent to the theoretical prediction from CFT and recent matrix product state (MPS) study~\cite{BernebigMPS}.
In the MPS work, they used the orbital cut to define the EE. It has a good scaling behavior in the region $L_y/l_B \in [10, 25]$ which is 
consistent to our real space EE study in Fig.~\ref{Fig3} as varying the aspect ratio in a finite size system. Moreover, our finite size calculation
can show the behavior of the EE for $L_y/l_B \rightarrow 0$ and $L_y/l_B \rightarrow \infty$ which have physical meaning of the Tao-Thouless crystal
and the edge-edge coupling respectively.  The MPS work did not look into these two regions because of the truncation errors.
For the RR state, although the system is limited by the numerical diagonalization, there are still strong evidences that  
the $e/5$ and $2e/5$ quasihole excitations have non-Abelian nature with quantum dimensions larger than one. 

With considering the effects of the disorder, we find that the TEE keeps invariant before the spectral gap closing which demonstrates that
the robustness of the topological properties. The spectral gap closing near $W_c \simeq 1.3$, at which the energy level statistics still satisfies 
the GOE distribution as shown in Fig.~\ref{Fig6} and Fig.~\ref{Fig7}.  As increasing the strength of the disorder, we find 
the average value of the adjacent gaps evolves from $r \simeq 0.53$ to $r \simeq 0.386$ 
which corresponds the GOE and Poisson distributions respectively. The critical strength of the disorder for MBL transition  is about $W \sim 10$.
It is larger than the $W_c$ since MBL phase need stronger disorder to localize all the eigenstates of the system.
The results of the energy level spacing statistics in these two limits are consistent
to the distributions. In the intermediate region of this MBL transition, we verify that the energy space follows
the distribution $P(x, \beta, \gamma_p)  \sim  x^{\beta}e^{-\eta x^{\gamma_p }}$. The behaviors of the parameter $\gamma_p$ 
and $\beta$ are similar to $r$.   

This work was supported by National Natural Science Foundation of China Grants No. 11674041, No. 91630205 and
Fundamental Research Funds for the Central Universities Grant No. CQDXWL-2014-Z006.

\end{document}